\documentstyle[preprint,aps]{revtex}
\begin{document}
\draft
\title{Higher Dimensional Reissner-Nordstr$\ddot{o}$m-FRW metric}
\author{Chang Jun Gao$^{1}$\thanks{E-mail: gaocj@mail.tsinghua.edu.cn}
Shuang Nan Zhang$^{1,2,3,4}$\thanks{E-mail:
zhangsn@mail.tsinghua.edu.cn}}
\address{$^{1}$Department of Physics and Center for Astrophysics, Tsinghua University, Beijing 100084, China(mailaddress)}
\address{$^2$Physics Department, University of Alabama in Huntsville, AL 35899, USA }
\address{$^3$Space Science Laboratory, NASA Marshall Space Flight Center, SD50, Huntsville, AL 35812, USA }
\address{$^4$Laboratory for Particle Astrophysics, Institute of High Energy Physics, Chinese Academy of Sciences, Beijing 100039, China}

\date{\today}
\maketitle
\begin{abstract}
\hspace*{7.5mm}By inspecting some known solutions to Einstein
equations, we present the metric of higher dimensional
Reissner-Nordstr$\ddot{o}$m black hole in the background of
Friedman-Robertson-Walker universe. Then we verify the solution
with a perfect fluid. The discussion of the event horizon of the
black hole reveals that the scale of the black hole would increase
with the expansion of the universe and decrease with the
contraction of the universe.
\end{abstract}
\section{introduction}
 \hspace*{7.5mm}Black holes in non-flat backgrounds are an important topic because
 astrophysical black holes are not asymptotically flat but embedded in our real
universe. In this respect, as early as in 1933, McVittie [1] found
his celebrated metric for a mass-particle in the FRW
(Friedman-Robertson-Walker) universe. It describes just the
Schwarzschild black hole which is embedded in the FRW universe
although there was no the notion of black hole at that time. In
1993, the multi-black-hole solution in the background of de Sitter
universe was discovered by Kastor and Traschen [2]. In 1999,
Shiromizu and Gen extended it to a spinning black hole [3]. In
2000, Nayak etc. [4, 5] studied the solutions for the
Schwarzschild and Kerr black holes in the background of the
Einstein universe. Recently, we extended the McVittie's solution to the charged case [6]. \\
\hspace*{7.5mm}On the other hand, with the development of string
theory, black holes in higher dimensional spacetimes have come to
play a fundamental role in physics. Furthermore, the possibility
of black hole production in high energy experiments has recently
been suggested in the context of the so-called TeV gravity. To
predict some observational and experimental results, we need
reliable knowledge about higher dimensional black holes. Exact
solutions for higher dimensional black holes have been constructed
by many authors [7-21].\\
\hspace*{7.5mm}Thus the aim of this paper is to extend our recent
work in which the McVittie solution is generalized to the
four-dimensional charged black hole to higher dimensions.
Spherically symmetric, vacuum, asymptotically flat spacetimes and
homogeneous, isotropic cosmological ones with fluid matter or
cosmological constant can be treated easily in general relativity
and give rise, respectively, to the Schwarzschild solution and the
FRW or de Sitter spacetimes. However, solutions representing an
isolated massive object, especially charged, embedded in an
expanding universe are much more difficult to obtain, and fully
explicit forms are not usually given (see, for example, the
general discussion in [22]). Therefore in this paper, we would
present the higher dimensional Reissner-Nordstrom-FRW metric by
simply inspecting some known solutions. Then we verify the
solution by substituting it into the Einstein-Maxwell equations
with a perfect
fluid. Finally we study the evolution of the event horizon of the black hole.\\
\section{higher dimensional Reissner-Nordstr$\ddot{o}$m-FRW metric }
\hspace*{7.5mm}The well-known four dimensional static
Schwarzschild metric in the isotropic spherical coordinates system
can be written as
\begin{equation}
ds^2=-\frac{\left(1-\frac{r_0}{x}\right)^2}{\left(1+\frac{r_0}{x}\right)^2}du^2
+{\left(1+\frac{r_0}{x}\right)^{{4}{}}}\left(dx^2+x^2d\Omega_{2}^2\right),
\end{equation}
where the constant $r_0$ is related to the mass of the black hole.
For our purpose, we write the McVittie solution which represents
the Schwarzschild-FRW metric as follows
\begin{eqnarray}
ds^2=-\frac{\left[\frac{{a^{\frac{1}{2}}}}{{\left(1+kx^2/4\right)^{\frac{1}{2}}}}-\frac{r_0}
{x{a^{\frac{1}{2}}}}\right]^2}{\left[\frac{{a^{\frac{1}{2}}}}{{\left(1+kx^2/4\right)
^{\frac{1}{2}}}}+\frac{r_0}{x{a^{\frac{1}{2}}}}\right]^2}du^2
+{\left[\frac{{a^{\frac{1}{2}}}}{{\left(1+kx^2/4\right)
^{\frac{1}{2}}}}+\frac{r_0}{x{a^{\frac{1}{2}}}}\right]^{{4}}}\left(dx^2+x^2d\Omega_{2}^2\right),
\end{eqnarray}
where $a=a(u)$ is the scale factor and $k$ is the curvature of the
universe. When $r_0=0$, it recovers the FRW metric. On the other
hand, when $a=const, k=0$, it is just the Schwarzschild metric.
For $k=0, a=e^{Hu}$, $H$ is a constant, Eq.(2) represents the
Schwarzschild-de Sitter metric. This form of Schwarzschild-de
Sitter metric can be reduced to the following familiar form via
coordinates transformations [6]
\begin{equation}
d{s}^2=-\left(1-\frac{r_0}{r}-H^2r^2\right)
d{t}^2+\left(1-\frac{r_0}{r}-H^2r^2\right)^{-1}d{r}^2+{r}^2d\Omega_{2}^2.
\end{equation}
\hspace*{7.5mm}Inspecting Eq.(1) and Eq.(2), we find that in order
to obtain the Schwarzschild-FRW metric, we need only to do the
following replacements in Eq.(1)
\begin{eqnarray}
&& 1\longrightarrow
\frac{{a^{\frac{1}{2}}}}{{\left(1+kx^2/4\right)^{\frac{1}{2}}}},\nonumber\\
&&\frac{r_0}{x}\longrightarrow \frac{r_0}{xa^{\frac{1}{2}}}.
\end{eqnarray}
Now we wonder whether the above method of replacement is
universal. So in the next let's
look for the higher dimensional version of McVittie solution.\\
\hspace*{7.5mm}For higher dimensional static Schwarzschild metric
\begin{equation}
ds^2=-\frac{\left(1^n-\frac{r_0^n}{x^n}\right)^2}{\left(1^n+\frac{r_0^n}{x^n}\right)^2}dt^2
+{\left(1^n+\frac{r_0^n}{x^n}\right)^{\frac{4}{n}}}\left(dx^2+x^2d\Omega_{n+1}^2\right).
\end{equation}
Make the above replacements in Eq.(5), then the higher dimensional
Schwarzschild-FRW metric is achieved
\begin{equation}
ds^2=-\frac{\left[\frac{{a^{\frac{n}{2}}}}{{\left(1+kx^2/4\right)^{\frac{n}{2}}}}-\frac{r_0^n}
{x^n{a^{\frac{n}{2}}}}\right]^2}{\left[\frac{{a^{\frac{n}{2}}}}{{\left(1+kx^2/4\right)
^{\frac{n}{2}}}}+\frac{r_0^n}{x^n{a^{\frac{n}{2}}}}\right]^2}dt^2
+{\left[\frac{{a^{\frac{n}{2}}}}{{\left(1+kx^2/4\right)
^{\frac{n}{2}}}}+\frac{r_0^n}{x^n{a^{\frac{n}{2}}}}\right]^{{\frac{4}{n}}}}\left(dx^2+x^2d\Omega_{n+1}^2\right).
\end{equation}
It is just the result we have obtained previously [20]. For the de
Sitter version, Eq.(6) can also be turned to our familiar form via
coordinates transformations
\begin{equation}
d{s}^2=-\left(1-\frac{r_0^n}{r^n}-H^2r^2\right)
d{t}^2+\left(1-\frac{r_0^n}{r^n}-H^2r^2\right)^{-1}d{r}^2+{r}^2d\Omega_{n+1}^2.
\end{equation}
So the method of replacement is likely universal. Let's look for
the four dimensional Reissner-Nordstr$\ddot{o}$m-FRW metric in
the following.\\
 \hspace*{7.5mm}For the four dimensional static
Reissner-Nordstr$\ddot{o}$m metric
\begin{equation}
ds^2=-\frac{\left[1^2-\frac{r_0^2}{x^2}+\frac{r_1^2}{x^2}\right]^2}{\left[\left(1+\frac{r_0}{x}\right)^2
-\frac{r_1^2}{x^2}\right]^2}du^2+\left[\left(1+\frac{r_0}{x}\right)^2
-\frac{r_1^2}{x^2}\right]^2\left(dx^2+x^2d\Omega_2^2\right),
\end{equation}
where the constant $r_1$ is related to the charge of the black
hole. Enlightened by Eq.(5), we make the following replacements
\begin{eqnarray}
&& 1\longrightarrow
\frac{{a^{\frac{1}{2}}}}{{\left(1+kx^2/4\right)^{\frac{1}{2}}}},\nonumber\\
&&\frac{r_0}{x}\longrightarrow \frac{r_0}{xa^{\frac{1}{2}}},
\nonumber\\
&&\frac{r_1}{x}\longrightarrow \frac{r_1}{xa^{\frac{1}{2}}},
\end{eqnarray}
then we obtain the four dimensional
Reissner-Nordstr$\ddot{o}$m-FRW metric
\begin{eqnarray}
ds^2&=&-\frac{\left[\frac{{a^{}}}{{\left(1+kx^2/4\right)^{}}}-\frac{r_0^2}{x^2a}+\frac{r_1^2}{x^2a}\right]^2}
{\left\{\left[\frac{{a^{\frac{1}{2}}}}{{\left(1+kx^2/4\right)^{\frac{1}{2}}}}+\frac{r_0}{xa^{\frac{1}{2}}}\right]^2
-\frac{r_1^2}{x^2a}\right\}^2}du^2\nonumber\\&&+\left\{\left[\frac{{a^{\frac{1}{2}}}}{{\left(1+kx^2/4\right)^{\frac{1}{2}}}}
+\frac{r_0}{xa^{\frac{1}{2}}}\right]^2
-\frac{r_1^2}{x^2a}\right\}^2\left(dx^2+x^2d\Omega_2^2\right).
\end{eqnarray}
It is just the result we have obtained [6]. For de Sitter version,
Eq. (10) can also be written in the Schwarzschild coordinates
\begin{equation}
d{s}^2=-\left(1-\frac{r_0}{r}+\frac{r_1^2}{r^2}-H^2r^2\right)
d{t}^2+\left(1-\frac{r_0}{r}+\frac{r_1^2}{r^2}-H^2r^2\right)^{-1}d{r}^2+{r}^2d\Omega_{2}^2.
\end{equation}
\hspace*{7.5mm}Now it seems that the method of replacement is
highly likely universal. We would admit it and conclude the higher
dimensional
Reissner-Nordstr$\ddot{o}$m-FRW metric with it.\\
\hspace*{7.5mm}The higher dimensional static
Reissner-Nordstr$\ddot{o}$m metric can be written as
\begin{eqnarray}
ds^2 =
-\frac{\left[1^{2n}-\frac{r_0^{2n}}{x^{2n}}+\frac{r_1^{2n}}{x^{2n}}\right]^2}
{\left[\left(1^n+\frac{r_0^n}{x^n} \right)^2
-\frac{r_1^{2n}}{x^{2n}}\right]^2}du^2
+\left[\left(1^n+\frac{r_0^n}{x^n}\right)^2
-\frac{r_1^{2n}}{x^{2n}}\right]^{\frac{2}{n}}\left(dx^2+x^2d\Omega_{n+1}^2\right).
\end{eqnarray}
Make the replacements in Eq.(9), then the higher dimensional
Reissner-Nordstr$\ddot{o}$m-FRW metric is obtained
\begin{eqnarray}
ds^2 &=&
-\frac{\left[\frac{{a^{{n}}}}{{\left(1+kx^2/4\right)^{{n}}}}-\frac{r_0^{2n}}
{x^{2n}a^{n}}+\frac{r_1^{2n}}{x^{2n}a^{n}}\right]^2}
{\left\{\left[\frac{{a^{\frac{n}{2}}}}{{\left(1+kx^2/4\right)^{\frac{n}{2}}}}
+\frac{r_0^n}{x^na^{\frac{n}{2}}} \right]^2
-\frac{r_1^{2n}}{x^{2n}a^{n}}\right\}^2}du^2
\nonumber\\&&+\left\{\left[\frac{{a^{\frac{n}{2}}}}{{\left(1+kx^2/4\right)^{\frac{n}{2}}}}
+\frac{r_0^n}{x^na^{\frac{n}{2}}} \right]^2
-\frac{r_1^{2n}}{x^{2n}a^{n}}\right\}^{\frac{2}{n}}\left(dx^2+x^2d\Omega_{n+1}^2\right),
\end{eqnarray}
namely
\begin{eqnarray}
ds^2 &=&
-\frac{\left[1-\frac{r_0^{2n}}{a^{2n}x^{2n}}\left(1+kx^2/4\right)^n+\frac{r_1^{2n}}{a^{2n}x^{2n}}
\left(1+kx^2/4\right)^n\right]^2}{\left\{\left[1+\frac{r_0^n}{a^nx^n}
\left(1+kx^2/4\right)^{\frac{n}{2}}\right]^2
-\frac{r_1^{2n}}{a^{2n}x^{2n}}\left(1+kx^2/4\right)^n\right\}^2}du^2\nonumber\\&
&
+\frac{a^2}{\left(1+kx^2/4\right)^2}\left\{\left[1+\frac{r_0^n}{a^nx^n}\left(1+kx^2/4\right)^{\frac{n}{2}}\right]^2
-\frac{r_1^{2n}}{a^{2n}x^{2n}}\left(1+kx^2/4\right)^{{n}}\right\}^{\frac{2}{n}}\nonumber\\&
&\cdot\left(dx^2+x^2d\Omega_{n+1}^2\right).
\end{eqnarray}
\hspace*{7.5mm}When $r_1=0$, the metric restores to the higher
dimensional McVittie solution. When $r_1=r_2=0$, it restores to
the higher dimensional FRW metric. When $a=const, k=0$, it
restores to the higher dimensional static
Reissner-Nordstr$\ddot{o}$m metric. When $a=const, k=1$, the
metric restores to the higher dimensional static
Reissner-Nordstr$\ddot{o}$m black hole in the Einstein universe.
In one word, it covers all the known solutions with the background
of FRW universe. For the de Sitter version, we show in the
following it can be reduced to our familiar form. To this end,
make variable transformation
\begin{equation}
 r=ax\left[\left(1+\frac{r_0^n}{a^nx^n}\right)^2-\frac{r_1^{2n}}{a^{2n}x^{2n}}\right]^{1/n},
\end{equation}
where $a=e^{Hu}$. Then Eq.(14) becomes
\begin{eqnarray}
d{s}^2
&=&-\left(1-\frac{4r_0^n}{{r^n}}+\frac{4r_1^{2n}}{{r}^{2n}}-H^2r^{2}\right)
d{u}^2-\frac{2Hr}{\sqrt{1-\frac{4r_0^n}{{r^n}}+\frac{4r_1^{2n}}{{r}^{2n}}}}dudr
\nonumber\\
&&+\left(1-\frac{4r_0^n}{{r^n}}
+\frac{4r_1^{2n}}{{r}^{2n}}\right)^{-1}d{r}^2+{r}^2d\Omega_{n+1}^2.
\end{eqnarray}
In order to eliminate the $dudr$ term, we introduce a new time
variable $t$, namely, $u\rightarrow t$
\begin{equation}
u=t-\int{\frac{Hr}{\left(1-\frac{4r_0^n}{{r^n}}+\frac{4r_1^{2n}}{{r}^{2n}}
-H^2r^{2}\right)\sqrt{1-\frac{4r_0^n}{{r^n}}+\frac{4r_1^{2n}}{{r}^{2n}}}}dr},
\end{equation}
Finally in the new coordinates system $(t,r)$, Eq.(16) is reduced
to
\begin{eqnarray}
d{s}^2
&=&-\left(1-\frac{4r_0^n}{{r^n}}+\frac{4r_1^{2n}}{{r}^{2n}}-H^2r^{2}\right)
d{t}^2
\nonumber\\
&&+\left(1-\frac{4r_0^n}{{r^n}}
+\frac{4r_1^{2n}}{{r}^{2n}}-H^2r^{2}\right)^{-1}d{r}^2+{r}^2d\Omega_{n+1}^2.
\end{eqnarray}
Absorb the constant $4$ by $r_0$ and $r_1$, we obtain the higher
dimensional Reissner-Nordstr$\ddot{o}$m-de Sitter metric in the
Schwarzschild coordinates system
\begin{eqnarray}
d{s}^2
&=&-\left(1-\frac{r_0^n}{{r^n}}+\frac{r_1^{2n}}{{r}^{2n}}-H^2r^{2}\right)
d{t}^2
\nonumber\\
&&+\left(1-\frac{r_0^n}{{r^n}}
+\frac{r_1^{2n}}{{r}^{2n}}-H^2r^{2}\right)^{-1}d{r}^2+{r}^2d\Omega_{n+1}^2.
\end{eqnarray}
\hspace*{7.5mm}In the next section, we will show our solution
Eq.(14) satisfies the Einstein-Maxwell equations.
\section{verification of the metric}
\hspace*{7.5mm}In the last section, we deduced the higher
dimensional Reissner-Nordstr$\ddot{o}$m-FRW metric. In this
section we will verify that it satisfies the Einstein-Maxwell
equations. The Einstein-Maxwell equations can be written as [23]
\begin{eqnarray}
G_{\mu\nu}&=&8 \pi \left(T_{\mu\nu}+E_{\mu\nu}\right), \nonumber
\\ F_{\mu\nu}&=& A_{\mu;\nu}-A_{\nu;\mu},\nonumber \\
F^{\mu\nu}_{;\nu}&=&4\pi J^{\mu}.
\end{eqnarray}
Here $J^{\mu}$ is the current density of the charge. $T_{\mu\nu}$
and $E_{\mu\nu}$ are the energy momentum for the perfect fluid and
electromagnetic fields, respectively, which are defined by
\begin{eqnarray}
T_{\mu\nu}&=&\left(\rho+p\right)U_{\mu}U_{\nu}+pg_{\mu\nu},
\nonumber \\
E_{\mu\nu}&=&\frac{1}{4\pi}\left(F_{\mu\alpha}F^{\alpha}_{\nu}-\frac{1}{4}g_{\mu\nu}F_{\alpha\beta}F^{\alpha\beta}\right),
\end{eqnarray}
where $\rho$ and $p$ are the energy density and pressure.
$U_{\mu}$ is the $(n+3)$-velocity of the particles. $F_{\mu\nu}$
and $A_{\mu}$ are the tensor and the potential
for electromagnetic fields.\\
\hspace*{7.5mm}Input the components of the metric Eq.(14) to the
Maple software package, we obtain the Einstein tensor $G_{\mu\nu}$
and then the energy momentum tensor $T_{\mu\nu}$ and $E_{\mu\nu}$,
respectively, for the perfect fluid and the electromagnetic fields
\begin{eqnarray}
&& T_0^0=-\rho, \ \ \ \ \ \ \ \ \ \ \ \
T_1^1=T_2^2=\cdot\cdot\cdot=T_{n+2}^{n+2}=p,\nonumber\\ &&8\pi
E_0^0=8\pi E_1^1=-8\pi E_2^2=-8\pi E_3^3=\cdot\cdot\cdot=-8\pi
E_{n+2}^{n+2}\nonumber\\ &&
=-\frac{2n\left(n+1\right)r_1^{2n}\left[1+kx^2/4\right]^n}{x^{2n+2}a^{2n+2}
\left\{\left[1+\frac{r_0^n}{a^nx^n}\left(1+kx^2/4\right)^{\frac{n}{2}}\right]^2
-\frac{r_1^{2n}}{a^{2n}x^{2n}}\left(1+kx^2/4\right)^n\right\}^{2+\frac{2}{n}}}.
\end{eqnarray}
Substituting the above components of electromagnetic tensor in the
second equation of Eqs.(21), we obtain the non-vanishing
components of electromagnetic tensor $F_{\mu\nu}$
\begin{eqnarray}
F^{01}&=&\frac{\sqrt{2n\left(n+1\right)}r_1^n\left(1+kx^2/4\right)^{1+\frac{n}{2}}}{x^{n+1}a^{n+2}
\left[1-\frac{r_0^{2n}}{a^{2n}x^{2n}}\left(1+kx^2/4\right)^{n}+\frac{r_1^{2n}}{a^{2n}x^{2n}}
\left(1+kx^2/4\right)^n\right]}\nonumber\\
&&
\cdot\frac{1}{\left\{\left[1+\frac{r_0^n}{a^nx^n}\left(1+kx^2/4\right)^{\frac{n}{2}}\right]^2
-\frac{r_1^{2n}}{a^{2n}x^{2n}}\left(1+kx^2/4\right)^n\right\}^{\frac{2}{{n}}}}.
\end{eqnarray}
Then substituting Eq.(23) in the second equation of Eqs.(20), we
obtain the non-vanishing components of the potential $A_{\mu}$
\begin{equation}
A_{0}=\int F^{01}g_{00}g_{11}dx.
\end{equation}
In the end, from the last equation of Eqs.(20), we obtain the
non-vanishing component of the flux density
\begin{eqnarray}
J^{0}&=&\frac{1}{4\pi\sqrt{-g}}\frac{\partial}{\partial x
}\left(\sqrt{-g}F^{10}\right)\nonumber\\
&
=&-\frac{1}{16\pi}kx^{-n}r_1^na^{-n-2}\left(n+2\right)\sqrt{2n\left(n+1\right)}\left(1+kx^2/4\right)^{n-n/2}\nonumber\\&&\cdot\left[1-\frac{r_0^{2n}}{a^{2n}x^{2n}}\left(1+kx^2/4\right)^{n}+\frac{r_1^{2n}}{a^{2n}x^{2n}}
\left(1+kx^2/4\right)^n\right]^{-1}\nonumber\\&&\cdot\left\{\left[1+\frac{r_0^n}{a^nx^n}\left(1+kx^2/4\right)^{\frac{n}{2}}\right]^2
-\frac{r_1^{2n}}{a^{2n}x^{2n}}\left(1+kx^2/4\right)^n\right\}^{-2/n},
\end{eqnarray}
where $g$ is the determinant of the metric tensor. We note that
$\sqrt{-g}F^{10}$ doe not depend on the variable $u$. It is the
function of only variable $x$. We also note that for the
space-flat universe, i.e., $k=0$, the flux density vanishes. For
$k\neq 0$, there is a charge density in the universe. The universe
of $k=0$ has the topology of $R^{3}$ and it is infinite both in
space and in radial variable $x$. The field lines of the charge
inside the black hole end in the infinity of the universe. So the
charge density is zero outside the black hole. On the other hand,
the universe of $k=1$ has the topology of $S^3$ and it is finite
in space. So there must be charge density in this universe to end
the field lines. For $k=-1$, the background universe is infinite
in space but the radial variable is finite. This is indicated by
the constraint of $1-kx^2/4\geq 0$ in Eq.(14). The field lines can
not
end in infinity. So charge density should also exist in this universe. \\
\hspace*{7.5mm}Up to now, we have verified the solution satisfies
the Einstein-Maxwell equations. Of course, any solution solves the
Einstein-Maxwell equations. However, not all the solutions are
physically meaningful. A solution is physically meaningful if and
only if it satisfies both the equations of fields and the
conditions of energy-momentum tensor. Fortunately, Eq.(14) meets
both the field equations Eqs.(20) and the energy-momentum
conditions Eqs.(21).\\
\section{event horizon of the black hole}
\hspace*{7.5mm}In this section, we make a discussion on the
evolution of the event horizon of the black hole. For simplicity
in mathematics, we consider the black holes in space-flat
universe. Set $k=0$ in the metric Eq.(14), we have
\begin{eqnarray}
ds^2 =
-\frac{\left(1-\frac{r_0^{2n}}{a^{2n}x^{2n}}+\frac{r_1^{2n}}{a^{2n}x^{2n}}
\right)^2}{\left[\left(1+\frac{r_0^n}{a^nx^n} \right)^2
-\frac{r_1^{2n}}{a^{2n}x^{2n}}\right]^2}du^2
+{a^2}{}\left[\left(1+\frac{r_0^n}{a^nx^n}\right)^2
-\frac{r_1^{2n}}{a^{2n}x^{2n}}\right]^{\frac{2}{n}}\left(dx^2+x^2d\Omega_{n+1}^2\right).
\end{eqnarray}
In order to study the event horizons of the black holes, we should
rewrite the metric in the Schwarzschild coordinates. So make
variable transformation
\begin{equation}
 r=ax\left[\left(1+\frac{r_0^n}{a^nx^n}\right)^2-\frac{r_1^{2n}}{a^{2n}x^{2n}}\right]^{1/n},
\end{equation}
Then Eq.(26) becomes
\begin{eqnarray}
d{s}^2
&=&-\left(1-\frac{r_0^n}{{r^n}}+\frac{r_1^{2n}}{{r}^{2n}}-H^2r^{2}\right)
d{u}^2-\frac{2Hr}{\sqrt{1-\frac{r_0^n}{{r^n}}+\frac{r_1^{2n}}{{r}^{2n}}}}dudr
\nonumber\\
&&+\left(1-\frac{r_0^n}{{r^n}}
+\frac{r_1^{2n}}{{r}^{2n}}\right)^{-1}d{r}^2+{r}^2d\Omega_{n+1}^2,
\end{eqnarray}
where $H\equiv \dot{a}/a$ which has the meaning of Hubble
parameter. Some constants have been absorbed by $r_0$ and $r_1$ in
Eq.(28). From the null surface equation,
\begin{equation}
g^{\mu\nu}\frac{\partial f}{\partial x^{\mu}}\frac{\partial
f}{\partial x^{\nu}}=0,
\end{equation}
where $f\equiv f\left(x^{\mu}\right)=0$ is the location of the
event horizon, we obtain the differential equation of evolution of
the event horizon $r_{EH}$
\begin{equation}
\frac{dr_{EH}}{du}=\left(-Hr_{EH}\pm\sqrt{1-r_0^n/r_{EH}^n+r_1^{2n}/r_{EH}^{2n}}\right)\sqrt{1-r_0^n/r_{EH}^n+r_1^{2n}/r_{EH}^{2n}}.
\end{equation}
We note that the two signs "$+$" and "$-$" in "$\pm$" are for
$H>0$ and $H<0$, respectively. Otherwise the cosmic event horizon
which is far away from the black hole, i.e., $r_{EH}\gg r_0$ and
$r_{EH}\gg r_2$, will expand or contract with the superluminal
motion, $|\dot{r}_{EH}|=|-Hr_{EH}\pm 1|>1$. This is physically
forbidden. Since the $g_{00}$ term in Eq.(28) is always negative
outside the black hole, we conclude
$\sqrt{1-r_0^n/r_{EH}^n+r_1^{2n}/r_{EH}^{2n}}>|Hr|$. So Eq.(30)
tells us the scale of the black hole would increase with the
expansion of the universe ($H>0$) and decrease with the
contraction of the universe ($H<0$). Compared with the four
dimensional black hole, it is easy to find that the
higher dimensional back hole would increase or decrease even faster. \\
\hspace*{7.5mm}On the other hand, Noerdlinger and Petrosion [24]
found that clusters or super-clusters would expand with the
expansion of the universe. Gautreau [25] also concluded that the
planetary orbits would expand by considering a model of a particle
embedded in an inhomogeneous, pressure free expanding universe.
Bonner [26] showed that a local system of electrically
counterpoised dust expands with the expansions of universe. Thus
our conclusion is consistent with their discussions.
\section{conclusion and discussion}
\hspace*{7.5mm}In conclusion, we have extended the four
dimensional Reissner-Nordstr$\ddot{o}$m-FRW metric to higher
dimensions. The solution covers all of the known metrics, such as
the higher dimensional static Reissner-Nordstr$\ddot{o}$m metric,
the higher dimensional static Reissner-Nordstr$\ddot{o}$m-de
Sitter metric, the McVittie metric and the four dimensional
Reissner-Nordstr$\ddot{o}$m-FRW metric. Then we verified the
solution by substituting it into the Einstein-Maxwell equations.
We find that there exists a charge density in the universe of
$k\neq 0$. It is due to the fact the field lines of charge inside
the black hole can not end in infinity in these two kinds of
universes. In the end, we make a discussion on the evolution of
the event horizon of the black hole. It is found that the scale of
the black hole would increase with the expansion of the universe
and decrease with the
contraction of the universe. This is consistent with the previous discussions.\\
\hspace*{7.5mm}\acknowledgements We thank the anonymous referee
for the expert and insightful comments, which have certainly
improved the paper significantly. This study is supported in part
by the Special Funds for Major State Basic Research Projects and
by the National Natural Science Foundation of China. SNZ also
acknowledges supports by NASA's Marshall Space Flight Center and
through NASA's Long Term Space Astrophysics Program.

\end{document}